\documentclass[sigconf]{acmart}

\usepackage{graphicx}
\usepackage{url} 
\usepackage{amsmath}
\usepackage{amsthm}
\usepackage{amsfonts}
\usepackage{bm}
\usepackage{caption}
\usepackage{multirow}
\usepackage[english]{babel}
\usepackage{subfigure}
\usepackage{booktabs}
\usepackage{array}
\usepackage{xcolor} 
\usepackage{pifont}
\usepackage{algorithmic}
\usepackage[linesnumbered,ruled,vlined]{algorithm2e}
\usepackage[most]{tcolorbox}

\AtBeginDocument{%
  }

\setcopyright{acmlicensed}
\copyrightyear{2026}
\acmYear{2026}
\acmDOI{XX}
\acmConference[Conference acronym 'XX]{Make sure to enter the correct
  conference title from your rights confirmation email}{June 03--05,
  2026}{xx, xx}

\begin{document}

\title{ScienceDB AI: An LLM-Driven Agentic Recommender System for Large-Scale Scientific Data Sharing Services}

\author{Qingqing Long, Haotian Chen, Chenyang Zhao, Xiaolei Du, \\ Xuezhi Wang,  Pengyao Wang, Chengzan Li,  Yuanchun Zhou\footnotemark[1], Hengshu Zhu}
\authornote{Corresponding Authors.}
\affiliation{%
  \institution{Computer Network Information Center, Chinese Academy of Sciences, Beijing, China}
   \country{}
}

\renewcommand{\shortauthors}{Qingqing Long et al.}

\begin{abstract}
The rapid growth of AI for Science (AI4S) has underscored the significance of scientific datasets, leading to the establishment of numerous national scientific data centers and sharing platforms. Despite this progress, efficiently promoting dataset sharing and utilization for scientific research remains challenging. Scientific datasets contain intricate domain-specific knowledge and contexts, rendering traditional collaborative filtering-based recommenders inadequate.
Recent advances in Large Language Models (LLMs) offer unprecedented opportunities to build conversational agents capable of deep semantic understanding and personalized recommendations. In response, we present \textbf{ScienceDB AI}, a novel LLM-driven agentic recommender system developed on Science Data Bank (ScienceDB), one of the largest global scientific data-sharing platforms. ScienceDB AI leverages natural language conversations and deep reasoning to accurately recommend datasets aligned with researchers' scientific intents and evolving requirements.
The system introduces several innovations: a Scientific Intention Perceptor to extract structured experimental elements from complicated queries, a Structured Memory Compressor to manage multi-turn dialogues effectively, and a Trustworthy Retrieval-Augmented Generation (Trustworthy RAG) framework. The Trustworthy RAG employs a two-stage retrieval mechanism and provides citable dataset references via Citable Scientific Task Record (CSTR) identifiers, enhancing recommendation trustworthiness and reproducibility.
Through extensive offline and online experiments using over 10 million real-world datasets, ScienceDB AI has demonstrated significant effectiveness, achieving more than 30\% improvement in offline metrics compared to advanced baselines and a over 200\% increase in click-through rates compared to keyword-based search engines. To our knowledge, ScienceDB AI is the first LLM-driven conversational recommender tailored explicitly for large-scale scientific dataset sharing services. The platform is publicly accessible at: https://ai.scidb.cn/en.
\end{abstract}


\maketitle

\section{Introduction}
\label{Introduction}

\begin{figure}[htbp]%
    \centering 
    \vspace{-2mm}    
     \subfigure[Yearly scientific dataset releases on Zenodo~\cite{sicilia2017community},  ScienceDB~\cite{zhou2024trusted} and FigShare~\cite{thelwall2016figshare}.]
     { \includegraphics[width=0.44\linewidth]{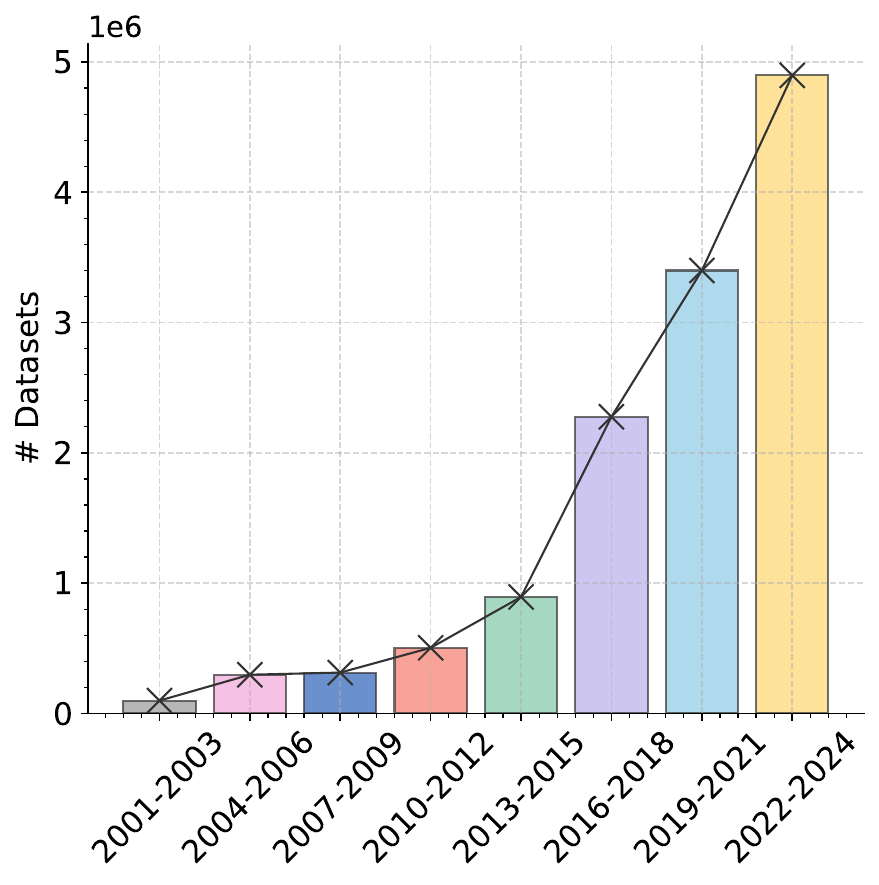} }
     \hspace{1mm}
     \subfigure[Dataset discipline distribution across different researcher behavior lengths in ScienceDB.]{ \includegraphics[width=0.44\linewidth]{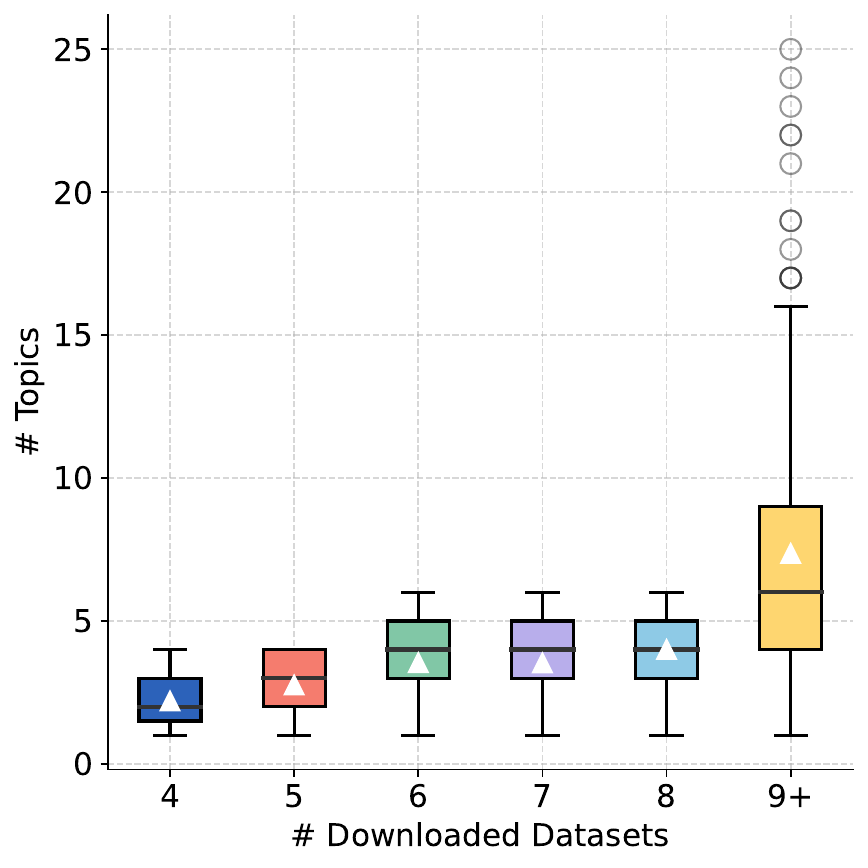} }
    \vspace{-2mm}
    \caption{Statistical results of datasets and user behaviors.} 
    \label{fig:intro}
    \vspace{-1mm}
\end{figure}

The rapid advancement of Artificial Intelligence for Science (AI4S)~\cite{camps2025artificial,kraemer2025artificial,luo2025large,zhu2025survey} has highlighted the critical importance of high-quality scientific data in accelerating discoveries across domains, including biology, physics, chemistry, and earth sciences ~\cite{kraemer2025artificial,messeri2024artificial,wang2024artificial,zhou2024artificial,sun2024nc}, etc. In response, governments and research institutions worldwide have established national scientific data centers~\cite{geer2010ncbi} and dataset-sharing platforms, such as the NCBI~\cite{geer2010ncbi}, OpenAIRE~\cite{rettberg2012openaire} and ScienceDB~\cite{zhou2024trusted}. These initiatives promote open access and foster collaborative use of scientific data, thereby enhancing its reusability. 
Consequently, the number of newly released scientific datasets has been significantly increasing in recent years~\cite{datafinder,thelwall2016figshare,sicilia2017community}, as illustrated in Fig. \ref{fig:intro} (a).

With the rapid growth of scientific datasets, enabling researchers to efficiently discover relevant datasets has become increasingly important. Effective dataset recommendation systems are therefore essential to facilitate  data-driven scientific discovery~\cite{datafinder,altaf2019dataset,ben2016dataset}. Traditional dataset recommenders generally fall into two categories. The first is \textit{behavior-based recommender}, which leverages user interaction histories through methods like Collaborative Filtering (CF)~\cite{agentcf,zhang2025unveiling,loveland2025understanding} and Graph Representation Learning (GRL)~\cite{chen2025lightgnn,liu2024selfgnn,ju2024comprehensive,long2021hgk}. The second is \textit{content-based recommender}, which rely on the query itself, including keyword-based retrieval~\cite{datafinder,zhou2024trusted} and semantic embedding-based matching~\cite{altaf2019dataset,ben2016dataset,long2025learning}. Existing dataset-sharing platforms, such as Google Dataset Search~\cite{brickley2019google}, DataCite Commons~\cite{ninkov2021datasets}, OpenAIRE~\cite{rettberg2012openaire} and Dryad~\cite{he2016reuse}, etc, all still rely heavily on keyword-based search engines. Their detailed information is shown in Table \ref{tab:compare_product}. While these works have achieved certain success, scientific dataset recommendation at scale introduces \textbf{unique challenges that are inadequately addressed}:

\begin{figure}[htbp]%
    \vspace{-1mm}
    \centering { \includegraphics[width=\linewidth]{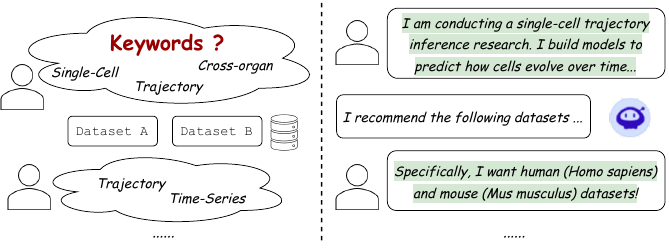} }
    \vspace{-5mm}
    \caption{The illustration of our motivation. The left figure shows the challenges of existing dataset sharing platforms. The right figure explains our ScienceDB AI can deeply understand the researcher's \textbf{experimental} dataset needs.} 
    \label{fig:intro_motivation}
\end{figure}

(1) Scientific dataset demands are often \textbf{task-specific and unrelated to historical behavior}. As illustrated in Fig. \ref{fig:intro} (b),  the x-axis denotes the number of datasets a researcher has previously downloaded in ScienceDB~\cite{zhou2024trusted}, while the y-axis indicates the number of distinct topics involved. Researchers with $\geq$9 downloads (about 10\% of the total) are grouped together. The figure reveals  weak topic consistency across a researcher's download history, implying that their dataset needs are driven more by evolving research tasks than persistent preferences. However, the user behavior-based recommenders are unsuitable in our scenario.

(2) Existing context-based recommenders \textbf{fall short in understanding experiment-level dataset needs}. Scientific exploration often involves highly specific, evolving, and nuanced dataset needs, expressed through rich natural language descriptions. Traditional keyword search or embedding-based matching falls short in understanding these complex requirements.
For instance, as depicted in Fig.\ref{fig:intro_motivation}, a researcher may query: "\textit{I am conducting a study on single-cell fate trajectory inference, focusing on cross-organ differentiation in human (Homo sapiens) and mouse (Mus musculus). I aim to build models that predict how individual cells evolve over time and respond to genetic perturbations}".  Such detailed and domain-specific intents require deep contextual understanding, which existing context-based recommenders are not equipped to handle effectively.

Fortunately, recent advances of LLMs and Agents in conversational recommendation offers a promising direction for addressing our problem~\cite{liang2024llm}. However, these models are inherently \textbf{prone to hallucination and forgetting issues}~\cite{guo2025deepseek,kraemer2025artificial}. They can generate hallucinated, non-existent, or inaccessible datasets. This poses a critical challenge in scientific scenarios, where trustworthy, accessibility, and citable are of the basic requirements~\cite{park2021reliable,farquhar2024detecting}.

In response, we propose the ScienceDB AI, an intelligent agentic recommender system designed for large-scale scientific data sharing service. Our system operates on a repository of over 10 million available datasets and introduces several key components to support  trustworthy, accessibility, and citable dataset recommendation. First, we develop a Experimental Intention Perceptor that extracts researchers’ data, topic, constraints, and evaluation criteria into a structured intent template. Second, we introduce a Structured Memory Compressor. It tracks user intent, dialogue context, and tool invocations in our multi-turn conversations, and summarize relevant historical information. This helps mitigate forgetting issues caused by the limited context window of LLMs. Third, to address the hallucination issues, we propose a Trustworthy Retrieval-Augmented Generation (Trustworthy RAG) framework. It incorporates a two-stage retriever to balance retrieval effectiveness and efficiency in our large-scale setting. To ensure dataset traceability and citation, we associate each dataset with a Citable Scientific Task Record (CSTR) and include direct links to CSTRs in the system’s responses. We conduct extensive offline and online evaluations in over 10 million real-world scientific datasets from ScienceDB platform. ScienceDB AI \textbf{achieves over a 30\% improvement in offline metrics} compared to existing agent-based recommenders. In \textbf{online A/B testing, it yields more than a 200\% increase} in Click-Through Rate (CTR) compared to traditional keyword-based search systems. We summarize our contributions as follows:

\begin{itemize}
    \item To the best of our knowledge, ScienceDB AI is the first LLM-driven agentic recommender system for a large-scale scientific data sharing services.
    \item We design a agentic framework, which consists of a experimental intention perceptor, a structured memory compressor, and a retriever-augmented recommender that attaches a CSTR to each dataset for trustworthy.
    \item Through extensive experiments over 10 million real-world datasets, ScienceDB AI achieves significant improvement (30\%+) in offline metrics, and remarkable increase (200\%+) in online A/B tests.
\end{itemize}

\section{Related Work}
In this section, we first review existing scientific dataset sharing platforms, highlighting their advantages and limitations. We then examine studies on dataset recommenders aimed at facilitating dataset discoverability. Finally, we discuss recent advances in agent-based conversational recommenders.

\begin{table*}[htbp]
\centering
\caption{Comparison of dataset sharing services.}
\vspace{-3mm}
\label{tab:compare_product}
\begin{tabular}{cccccccc}
\toprule
\textbf{Product/Platform} & \textbf{\# Disciplines} & \textbf{For Research} & \textbf{Sharing SourceData} & \textbf{\# Datasets} & \textbf{CRS}  \\
\midrule
    DataCite Commons~\cite{ninkov2021datasets} & >10 & \textcolor{green}{\ding{51}} & \textcolor{green}{\ding{51}} & 42,896,080 &  \textcolor{red}{\ding{55}}\\  
    Google Dataset Search~\cite{brickley2019google} &  >10 & \textcolor{green}{\ding{51}} & \textcolor{red}{\ding{55}} (Only Metadata)  & 25 Million &  \textcolor{red}{\ding{55}}\\  
    Zenodo~\cite{sicilia2017community} & <5 & \textcolor{green}{\ding{51}} & \textcolor{green}{\ding{51}} & 4 Million &  \textcolor{red}{\ding{55}}\\
    OpenAIRE~\cite{rettberg2012openaire} & >10 & \textcolor{green}{\ding{51}} & \textcolor{green}{\ding{51}}  & 8,382,956 &  \textcolor{red}{\ding{55}} \\
    PaddlePaddle~\cite{paddle}  &  >10 & \textcolor{green}{\ding{51}} & \textcolor{green}{\ding{51}}  & $\sim$10,000 &  \textcolor{red}{\ding{55}}\\ 
    Dataverse~\cite{magazine2011dataverse} & >10  & \textcolor{green}{\ding{51}} & \textcolor{green}{\ding{51}}  & 139,231 &  \textcolor{red}{\ding{55}}\\  
    CKAN~\cite{winn2013open} & >10  & \textcolor{green}{\ding{51}} & \textcolor{green}{\ding{51}}  & 24,233 &  \textcolor{red}{\ding{55}}\\ 
    Dryad~\cite{he2016reuse} & <5  & \textcolor{green}{\ding{51}} & \textcolor{green}{\ding{51}}  & $\sim$900,000 &  \textcolor{red}{\ding{55}}\\ 
    Snowflake Marketplace~\cite{bell2021data} & Unknown  &  \textcolor{red}{\ding{55}} (Commercial) & \textcolor{green}{\ding{51}}   & Unknown &  \textcolor{red}{\ding{55}} \\ 
    DataBricks~\cite{l2022databricks} &  Unknown &  \textcolor{red}{\ding{55}} (Commercial) & \textcolor{green}{\ding{51}}   & Unknown &  \textcolor{red}{\ding{55}}\\ 
    HuggingFace~\cite{jain2022hugging} & <3 & \textcolor{green}{\ding{51}} & \textcolor{green}{\ding{51}} & 461,199 &  \textcolor{red}{\ding{55}} \\
    RADx Data Hub~\cite{martinez2025radx} & <2 & \textcolor{green}{\ding{51}} & \textcolor{green}{\ding{51}}  & $\sim$5,000 &  \textcolor{red}{\ding{55}}\\  
    NCBI~\cite{geer2010ncbi} & <3 & \textcolor{green}{\ding{51}} & \textcolor{green}{\ding{51}}  & $\sim$1,000 &  \textcolor{red}{\ding{55}}\\ 
    FigShare~\cite{thelwall2016figshare} & <3 & \textcolor{green}{\ding{51}} & \textcolor{green}{\ding{51}} & $\sim$380,000 &  \textcolor{red}{\ding{55}}\\
    \midrule
    \textbf{ScienceDB AI (Ours)} & \textbf{All (>18)}  & \textcolor{green}{\ding{51}} & \textcolor{green}{\ding{51}}  & \textbf{10 Million} & \textcolor{green}{\ding{51}} \\ 
\bottomrule
\end{tabular}
\end{table*}

\subsection{Scientific Dataset Sharing Platforms}
The recent advancement of AI4S has shown the critical importance of high-quality scientific data~\cite{wang2024artificial,sun2024nc}. Governments and research institutions worldwide have established national scientific data centers and dataset-sharing platforms. Here we compare 14 existing dataset sharing platforms across five dimensions: (1) the number of supported disciplines, (2) whether they are designed for research use cases, (3) whether they provide source data, (4) the number of available datasets, and (5) the presence of Conversational Recommendation Systems (CRS). The number of disciplines is estimated based on the primary discipline taxonomy of OpenAlex~\cite{bordignon2024openalex}. A detailed comparison is provided in Table \ref{tab:compare_product}. Snowflake Marketplace~\cite{bell2021data} and  DataBricks~\cite{l2022databricks} are two commercial products, thus their dataset information is unknown. As shown in the table, half of the platforms support around 10 disciplines, while the rest support fewer than five. In contrast, our platform covers 18 first-level disciplines, providing broader subject coverage and more diverse, domain-specific datasets. Among all platforms, Google Dataset Search~\cite{brickley2019google}, ScienceDB~\cite{zhou2024trusted} and DataCite Commons~\cite{ninkov2021datasets} host the largest number of datasets. However, Google Dataset Search only indexes metadata without providing source data, limiting its applicability for experimental research. 

In summary, existing data platforms lack effective support for dataset sharing and recommendation. In contrast, ScienceDB AI stands out as the only data center that enables intelligent recommendations, allowing researchers to express complex data needs in natural language and efficiently discover relevant datasets, which ultimately accelerate scientific discovery.

\subsection{Dataset Recommenders}
Recent years there are only three representative works designed for the dataset recommendation task.
DataFinder~\cite{datafinder} proposes a text similarity based dataset recommendation model. It takes BERT as the embedding model for dataset description and the user's input query. Altaf et al.~\cite{altaf2019dataset} propose a variational graph autoencoder for query-based dataset recommendation tasks. It construct a set of research papers, which reflects a user's research interest. The recommended datasets are based on the representation similarity of the dataset description and the constructed graph of research papers for the user. DataLinking~\cite{ben2016dataset} uses concept frequency and TF-IDF to extract the similarity features of user queries and dataset descriptions. 
However, all these works are primarily keyword-based and cannot understand the researchers' complex needs or support interactive, natural language-based queries.

\subsection{Agent-based Conversational Recommenders}
Sorts of studies have shown~\cite{Macrec,yan2025agentsociety} LLM and Agent-based conversational recommendation systems have the better performance of understanding user's complicated intentions than traditional models. They have the ability to leverage specialized tools, which can relieve the limited knowledge due to model scale and pretrained data size constraints. 
Representative works include AgentCF~\cite{agentcf}, InteRecAgent~\cite{InteRecAgent} and CoSearchAgent~\cite{Cosearchagent}, etc. Specifically, CoSearchAgent, Fang et al.~\cite{fang2024multi}, and MACRec~\cite{Macrec} are multi-agent collaborative search systems. However, the multi-agent system has communication delays, which brings longer system response time, further can not suit well for a large-scale online recommendation scenario. Thus this work pay attention to the single-agent recommendation works. AgentCF designs agent-based collaborative filtering to simulate user-item interactions. InteRecAgent, ChatCRS~\cite{ChatCRS}, and RecMind~\cite{RecMind} design agent-based conversational frameworks, which contains mechanisms of planning, memory, web search, reflection and recommendation tools. Other agent-based works~\cite{wei2025enhanced,guo2024knowledge,zhao2024let,Instructagent} mainly focus on personalized recommendations in conversations. 

However, all the above models are inherently prone to hallucination~\cite{kraemer2025artificial}, often generating recommendations for non-existent or inaccessible datasets. This presents a critical challenge in scientific settings, where trustworthy, accessibility, and citable are of the basic requirements~\cite{park2021reliable,farquhar2024detecting}. Moreover, these models are primarily behavior-based models, and thus unsuitable for understanding experiment-level queries.

\begin{figure*}[t]
    \centering
     \vspace{-6mm}
    \includegraphics[width=\linewidth]{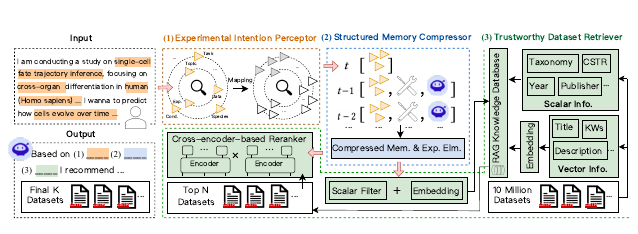}
    \vspace{-8mm}
    \caption{Technical framework of our designed ScienceDB AI system. It consists of experimental intention perceptor, structured memory compressor, and a retriever-augmented recommender that attaches the CSTR~\cite{zhou2024trusted} to each dataset for trustworthiness.}
    \label{fig:framework}
\end{figure*}

\section{Technical Details of ScienceDB AI}
In this section, we provide the technical detailed of ScienceDB AI. First, we  provide a overview of our technical framework and problem definition. Then we introduce our framework components, i.e., Experimental Intention Perceptor, Structured Memory Compressor and a retriever-augmented recommender that attaches a unique identifier to each dataset for trustworthy.

\subsection{Framework Overview}
\begin{figure*}[htbp]%
    \centering
    \subfigure[The Entrance]{ \includegraphics[width=0.315\linewidth]{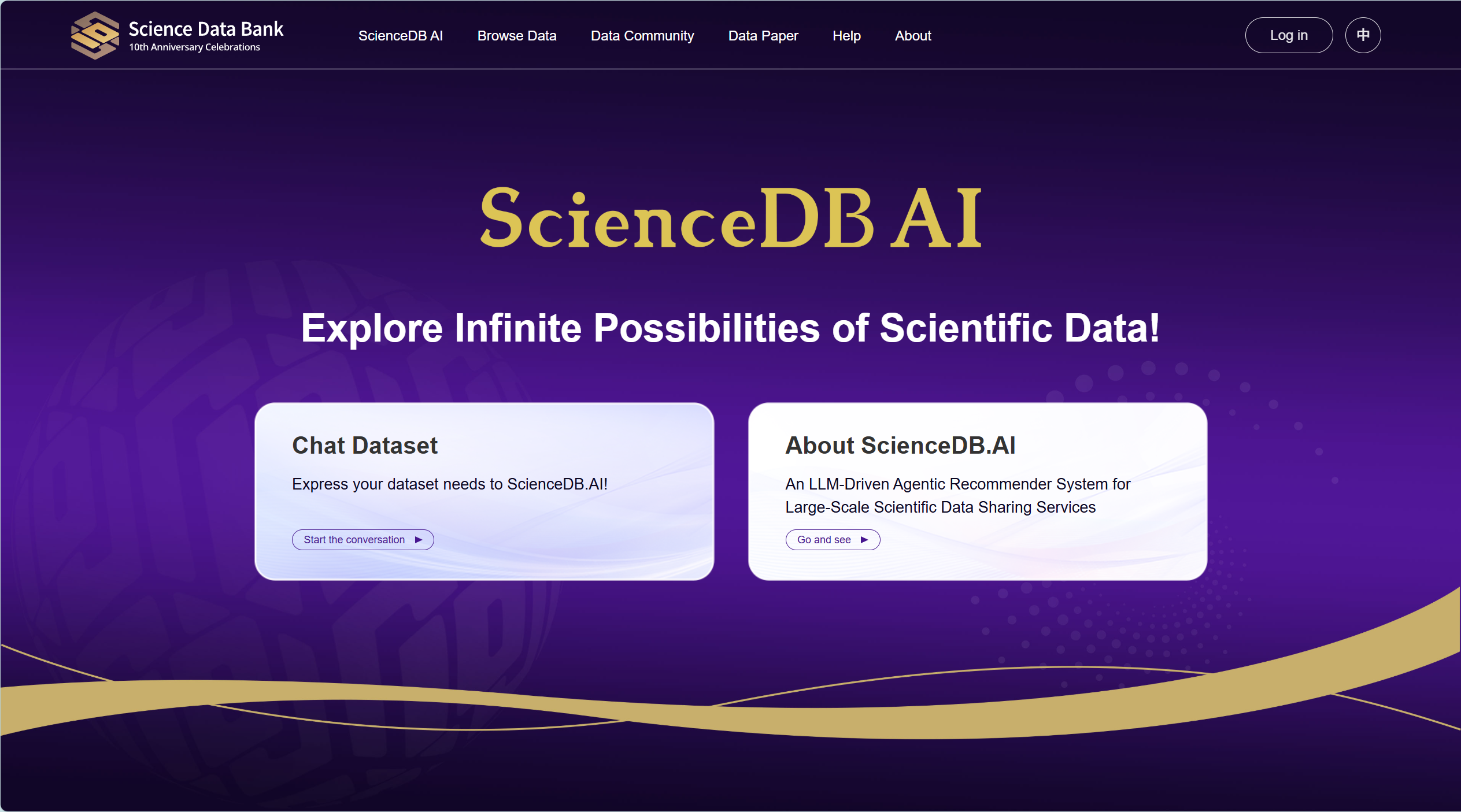} }
    \hspace{2mm}
    \subfigure[A Chat Example]{\includegraphics[width=0.30\linewidth]{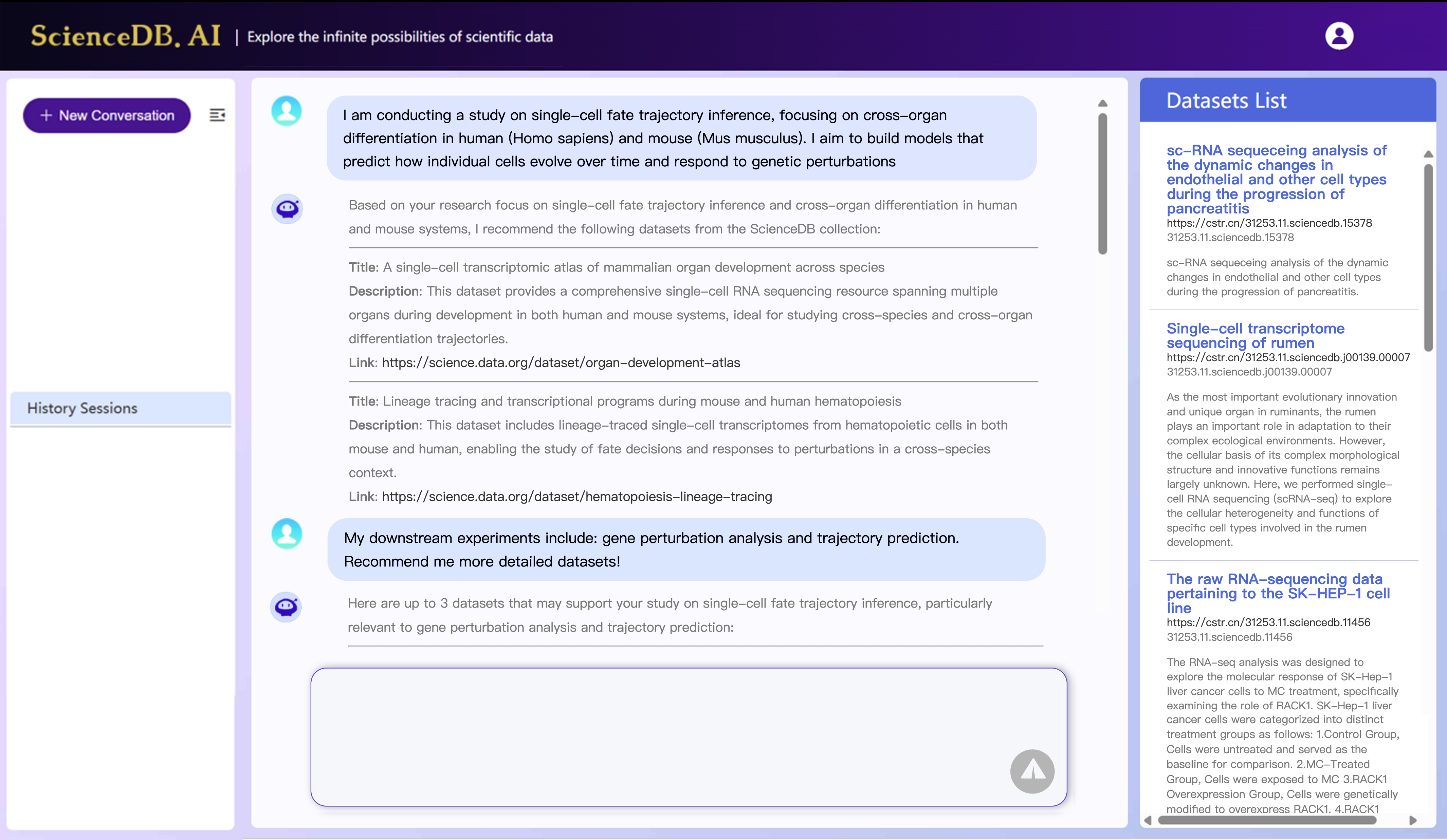} }
    \hspace{2mm}
    \subfigure[The 10 Million Dataset Example]{\includegraphics[width=0.30\linewidth]{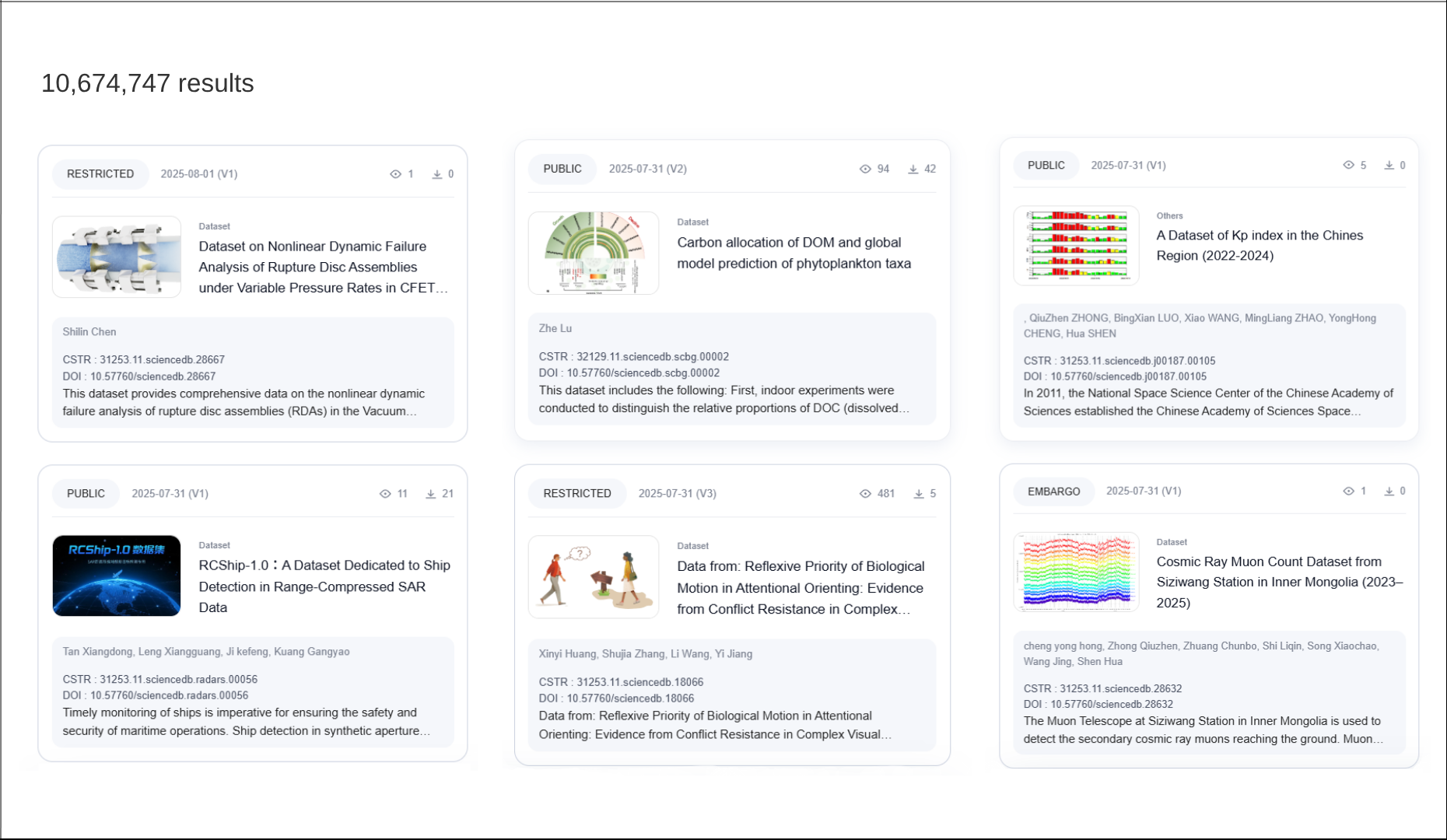} }
    \vspace{-2mm}
    \caption{Our online ScienceDB AI platform, which can be visited at  \url{https://ai.scidb.cn/en}.}
    \label{fig:online_system}
\end{figure*}

\textbf{Framework Overview}. 
The overall technical framework of ScienceDB AI is shown in Fig.~\ref{fig:framework}, which consists of a experimental intention perceptor, a structured memory compressor, and a retriever-augmented recommender that attaches  a unique identifier to each dataset for trustworthiness. Our online ScienceDB AI system can be visited at \url{https://www.ai.scidb.cn/en}. Our online web examples are shown in Fig.~\ref{fig:online_system}. 

\noindent\textbf{Problem Definition}. Let $\bm{Q} = \{\bm{q}_1,\bm{q}_2,...,\bm{q}_T\}$ denote a multi-turn researcher's query, where $\bm{q}_t$ denotes the $t$-th turn input query, which uses technical descriptions and contains research goals, methodological descriptions, experimental constraints, etc. Let $\bm{D} = \{\bm{d}_1,\bm{d}_2,...,\bm{d}_N\}$ denote the large-scale candidate datasets, where $N$ is larger than 10 million in this paper. Each dataset $\bm{d}_i$ has the corresponding metadata information and a textual description. This paper aims at designing a dataset recommender $\mathcal{F}$, which recommends the most suitable $K$ $(K \ll N)$ datasets for researchers with as few conversations as possible, i.e, making $T$ as small as possible.

\subsection{Experimental Intention Perceptor}
\label{sec:model_2}
As shown in Fig.\ref{fig:intro_motivation} and Fig.\ref{fig:dataset-sample-1}, the experimental inputs of researchers can be extremely complicated. To support experiment-level dataset recommendation for researchers, we design an Experimental Intention Perceptor that extracts a researcher’s long-passage natural language into \textbf{structured experimental elements}. Compared with traditional dataset recommendation models~\cite{datafinder,altaf2019dataset,ben2016dataset} and general recommenders~\cite{InteRecAgent,Cosearchagent}, this paper aims at a conversational dataset recommender, which is specially designed for scientific research scenarios.

The intention perceptor is designed based on the structured element system and typical process  of scientific discovery~\cite{martin1998elements,mccomas1998principal,peffers2007design}. Specifically, \textit{Data}, \textit{Topic}, \textit{Experimental Constraints/Settings}, and \textit{Evaluation Metrics} are typical top-level elements. The \textit{Species} and \textit{Data Modality}, \textit{Source}, and \textit{Annotation} are typical second-level elements of \textit{Data}. 
Take the input query in Fig.~\ref{fig:framework} as example,  our intention perceptor identifies the research topic as cross-organ cell differentiation in human, the task as single-cell fate trajectory inference and cells evolve over time. The cross-organ scope and human tissue context are interpreted as experimental constraints. The extracted scientific intention of a query  will be rewritten as $\Tilde{\bm{q}_t}$.

\subsection{Structured Memory Compressor}
Due to the complexity of researchers' needs, their requests can be lengthy and often require more rounds of conversations compared with general recommendation tasks. To effectively support multi-turn, complicated queries in scientific scenarios, we design a Structured Memory Compressor that distills essential information from a long dialogue history while preserving context-dependent dependencies. This module addresses the challenges of inherently forgetting issues~\cite{guo2025deepseek,kraemer2025artificial} of LLMs.

We track all the real-time dialogue states and histories in our platform. Let $\Theta_{1:T}$ denote the dialogue history up to turn $t$, then
\begin{equation}
    \Theta_{1:T} = \{(\Tilde{\bm{q}_1}, \bm{\tau}_1, \bm{r}_1), \dots, (\Tilde{\bm{q}_T}, \bm{\tau}_T, \bm{r}_T)\},
\end{equation}
where $\bm{\tau}_t$ represents the tool calling and execution logs. The tool logs are able to avoid redundant operations in the next turns of conversation. $\bm{r}_t$ denotes the response of our ScienceDB AI at turn $t$. 
The memory budget is limited to $L_{\text{max}}$ tokens (e.g., 32K), and thus full inclusion of $\Theta_{1:T}$ is meaningful and challenging. If an extremely long conversation record is directly input into LLMs, it will cause the LLM to forget the system prompt or the given set of recommended candidate datasets, thereby leading to hallucinations in the response.
In this paper, we aim to compress $\Theta_{1:t}$ into a structured memory $\mathcal{S}_t$ that retains information in the previous $t-1$ turns,
\begin{equation}
    \mathcal{S}_t = \begin{cases}
    \Theta_{1:1} , & t=1. \\
    \mathcal{M}(\Tilde{\bm{q}_t}, \bm{\tau}_t, r_t, \mathcal{S}_{t-1}), & t>1.
    \end{cases}
\end{equation}
Besides, $\mathcal{S}_t$ is expected to be recency-aware conflict resolution, which prefers recent updates over stale or outdated ones. Here we conduct explicit compression, rather than implicit compression~\cite{dai2025pretraining} for maintaining the structured intention template. $\mathcal{M}$ denotes a LLM-based Agent to summarize the historical conversational logs into structured information.
Then compressed structured memory $\mathcal{S}_t$ is taken as the context for the final response of LLMs. When conflicts are unresolved due to semantic ambiguity, we proactively generate a clarification question, such as "Do you want to override your previous dataset constraint ...?"

\subsection{Trustworthy Dataset Retriever}
\label{sec:model_Retriever}
To enable more accurate retrieval candidates, we adopt a \textit{two-stage} retriever for the trade-off between effectiveness and efficiency in our large-scale dataset sharing service. Each dataset $\bm{d}_i$ is associated with both dense embeddings and structured metadata, such as publication time and affiliated institution (as shown in Fig.\ref{fig:qdrant-sample} in the Appendix).
In the first stage, we retrieve top-$N$ candidate datasets using vector similarity with pre-filtering. If the input query explicitly includes or an LLM extracts, constraints such as publication date, taxonomy, or affiliated institution, we apply scalar filtering to reduce the candidate space. We then compute the cosine similarity between the query embedding $\bm{e}(\Tilde{q_t})$ and dataset descriptions $\bm{e}(d_i)$ and select the top-$N$ most similar datasets. In the second stage, we aims at deeply understanding a researcher's intention, we then adopt ColBert~\cite{Colbert} as the reranker. 
The reranker performs fine-grained late interaction between the token-level embeddings of  $\Tilde{\bm{q}_t}$  and $N$ candidates, and produces a final top-$K$ datasets. Note that the number of recommended dataset in the final response is based on the researcher's needs. If not specifically specified in the input query, the $K$ is set to 3.

Furthermore, to ensure that the recommended datasets are both traceable and trustworthy, i.e., uniquely identifiable and citable, we attach a \textit{Citable Scientific Task Record (CSTR)}~\cite{zhou2024trusted} to each dataset $d_i$ and include the corresponding CSTR links in our final response. The CSTR identifier provides a unique and standardized ID for scientific resources, similar to a DOI~\cite{li2022tracing}. However, CSTR supports a wider range of resource types. In our scenario, it can uniquely identify both the dataset and its source data files, while the DOI cannot. To be specific, it  helps eliminate ambiguity caused by changes in names or storage locations of the dataset and its source files. To enforce this behavior, we incorporate a \textit{system prompt} as: "\textit{For each selected dataset, you MUST return its CSTR identification}."

The pseudocode of our technical framework of ScienceDB AI is shown in Algorithm~\ref{alg:main}.

\begin{algorithm}[ht]
\small
\caption{Algorithm workflow of our ScienceDB AI.}
\label{alg:main}
\KwIn{User query $\bm{q}_t$ at turn $t$; Dialogue history $\Theta_{1:t-1}$; Dataset index $\mathcal{D}$ with metadata}
\KwOut{Top-$k$ recommended datasets $\{d_1, \dots, d_k\}$ and final response $r_t$}
\BlankLine
\textbf{Initialize:} Structured memory $\mathcal{S}_0 \leftarrow \varnothing$

\BlankLine
\textbf{Step 1: Experimental Intention Perceptor}\;
$\Tilde{\bm{q}_t} \leftarrow \textsc{LLMParse}(\bm{q}_t, \Theta_{1:t-1})$ 
\tcp*[l]{Parse query and dialogue history with LLM to extract scientific intention}
Decompose $\Tilde{\bm{q}_t} = (\mathcal{U}, \mathcal{T}, \mathcal{D}, \mathcal{E}, \mathcal{Z})$
\tcp*[l]{Subject $\mathcal{U}$, Task $\mathcal{T}$, Data Modality $\mathcal{D}$, Experimental Settings $\mathcal{E}$, Evaluation Metrics $\mathcal{Z}$}

\BlankLine
\textbf{Step 2: Structured Memory Compressor}\;
Update dialogue logs: $\Theta_{1:t} \leftarrow \Theta_{1:t-1} \cup \{(\Tilde{\bm{q}_t}, \bm{\tau}_t, \bm{r}_t)\}$\;
Compress $\Theta_{1:t}$ into structured memory: $\mathcal{S}_t \leftarrow \textsc{SSRC}(\Theta_{1:t})$
\tcp*[l]{Scientific Semantic Retention Compression (SSRC) to summarize history into structured memory}

\BlankLine
\textbf{Step 3: Trustworthy Dataset Retriever}\;
Embed intent: $\mathbf{h}_t \leftarrow \textsc{EmbedIntent}(\Tilde{\bm{q}_t}, \mathcal{S}_t)$\;
Embed datasets: $\mathbf{H}_\mathcal{D} = \{\mathbf{h}_d \mid d \in \mathcal{D}\}$
\tcp*[l]{Each $\mathbf{h}_d$ encodes dataset metadata: description, keywords, source, etc.}
Retrieve top-$k$ candidates via approximate nearest neighbor (ANN):
\[
\{d_1, \dots, d_k\} \leftarrow \textsc{ANN}(\mathbf{h}_t, \mathbf{H}_\mathcal{D})
\]

Re-rank candidates via cross-encoder:
\[
\text{score}(\Tilde{\bm{q}_t}, d_i) \leftarrow f_{\text{cross}}(\Tilde{\bm{q}_t}, \text{meta}(d_i))
\]

\BlankLine
\textbf{Step 4: Generate Final Response}\;

\[
r_t \leftarrow \textsc{LLMAnswer}(\Tilde{\bm{q}_t}, \mathcal{S}_t, \{(d_i, \text{meta}(d_i))\}_{i=1}^k, \text{SystemPrompt})
\]

\BlankLine
\Return{$\{d_1, \dots, d_k\},r_t$}
\end{algorithm}

\paragraph{\textbf{Discussion}.}  
Compared with other LLM or Agent-based recommendation models, we show that the researcher intent understanding, retriever, and memory modules have the most significant impact on meeting researchers’ scientific needs in large-scale data sharing service, more so than complex planning, web search, or reflection modules. Experimental evidence supporting this claim is provided in Section \ref{sec:exp_overall}.

\begin{table*}[t]
\centering
\caption{The overall performance comparision in multi-turn conversational dataset recommendation.}
\label{tab:main_res}
\begin{tabular}{ccccccccccccccccc}
\toprule
\multirow{2}{*}{\textbf{Model}} & \multicolumn{3}{c}{\textbf{Recall}} & \multicolumn{3}{c}{\textbf{NDCG}} & \multicolumn{3}{c}{\textbf{MRR}}  &  \multicolumn{3}{c}{\textbf{AT}} \\
\cmidrule(lr){2-4}   \cmidrule(lr){5-7}  \cmidrule(lr){8-10}   \cmidrule(lr){11-13} 
 & @1 & @3  & @5 &  @1 & @3  & @5 &  @1 & @3  & @5 & @1 & @3  & @5 \\
\midrule
DataFinder~\cite{datafinder} & 0.0115 & 0.0726 & 0.1481 & 0.0115 & 0.0455 & 0.0764  & 0.0115 & 0.0363 & 0.0533 & 3.35 & 3.09 & 3.01 \\
DataLinking~\cite{ben2016dataset} & 0.2605 & 0.3003 & 0.3084 & 0.2605 & 0.2838 & 0.2871 & 0.2605 & 0.2781 & 0.2800  & 3.23 & 3.06 & 3.03  \\
\midrule
DeepSeek-V3~\cite{deepseek}+RAG & 0.2277  & 0.2513  & 0.2530 & 0.2277 & 0.2420  & 0.2428  & 0.2277  & 0.2388  & 0.2392 & 3.33 & 3.24 & 3.23 \\ 
Qwen3~\cite{qwen3}+RAG & 0.2559 & 0.2778 & 0.2824  & 0.2559 & 0.2692 & 0.2712 & 0.2559 & 0.2662 & 0.2673 & 3.21 & 3.17 & 3.15 &  \\  
\midrule
InteRecAgent~\cite{InteRecAgent} & 0.2686 & 0.3083 & 0.3141 & 0.2686 & 0.2926 & 0.2950 & 0.2684 & 0.2871 & 0.2884 & 3.20 & 3.06 & 3.05  \\  
CoSearchAgent~\cite{Cosearchagent} & 0.1608 & 0.1988 & 0.2386 & 0.1608 & 0.1822 & 0.1984 & 0.1608 & 0.1766 & 0.1854 & 3.41 & 3.31 & 3.25 \\ 
\midrule
\textbf{Ours} & \textbf{0.4064} & \textbf{0.4187} & \textbf{0.4196} & \textbf{0.4065} & \textbf{0.4142} & \textbf{0.4146} & \textbf{0.4065} & \textbf{0.4126} &  \textbf{0.4128} & \textbf{3.19} & \textbf{2.89} & \textbf{2.83}\\ \bottomrule
\end{tabular}
\end{table*}

\section{Experiment}
In this section, we first introduce the experimental settings used to evaluate our approach. Then, we present the overall performance results and analyze the running efficiency of ScienceDB AI. Subsequently, we provide a detailed case study to illustrate practical effectiveness. 
Finally, we report results from an online A/B test to comprehensively validate our framework.

\subsection{Experimental Settings}
\paragraph{\textbf{Dataset}}
We construct our offline evaluation dataset by randomly sampling user-dataset click logs from ScienceDB~\cite{zhou2024trusted,li2022tracing} over the past two years. Specifically, we sample approximately 10,000 users and 15,000 corresponding downloaded datasets. For each researcher, \textbf{the dataset they previously clicked is treated as the ground-truth target in the simulated conversation}. Candidate datasets are retrieved from 10 million datasets based on cosine similarity.
Following previous conversational recommendation works~\cite{liang2024llm,InteRecAgent}, we construct an offline dataset with multi-turn interactions to simulate the complex and professional needs of researchers. To better simulate these complexities, we leverage a LLM (Qwen-Plus) to generate experimental design plans based on dataset descriptions. Compared with existing offline conversational datasets,  our input queries are significantly more detailed, lengthy, and nuanced, posing a more challenging conversational recommendation task. The conversation turn is set between 3 to 5. The detailed offline constructed process and samples are shown in Section~\label{sec:Appendix_Offline_Dataset} \textbf{in the Appendix}.

\paragraph{\textbf{Competitive Baselines}}
We select the following baselines as our competitors, which can be classified into three categories:
(1) \textit{Dataset Recommenders}. DataLinking~\cite{ben2016dataset} and DataFinder~\cite{datafinder}. DataLinking uses concept frequency and TF-IDF to extract the similarity features of user query and dataset descriptions. DataFinder proposes a text similarity based dataset recommendation model, which takes BERT as the embedding model for dataset descriptions. (2) \textit{Dialogue Recommenders}. DeepSeek-V3:671b~\cite{deepseek} (2025-03-24) and Qwen3:235B~\cite{qwen3}. (3) \textit{Agent-based Conversational Recommenders}. CoSearchAgent~\cite{Cosearchagent} and InteRecAgent~\cite{InteRecAgent}. 
CoSearchAgent is a multi-agent collaborative system that effectively supports multi-user conversations.

\paragraph{\textbf{Evaluation Metrics}}
Following previous works~\cite{datafinder,ben2016dataset}, we use popularly used recommendation metrics, i.e., top-$K$ Recall, Normalized Discounted Cumulative Gain (NDCG) and Mean Reciprocal Rank (MRR), as our offline evaluators. As this paper focus on accurate recommendation towards scientific scenarios, we focus on the @1, @3 and @5 of the above metrics. The detailed offline metric information is shown in \ref{appendix:llm-metrics}  in the Appendix. We also adopt the Average Turns (AT) required for a successful recommendation in our multi-turn conversations. 
Unsuccessful recommendations within $t$ rounds are recorded as $t+1$ in calculating AT. For online performance evaluation, we consistently take the Click-Through-Rate (CTR) as the primary metric.

\paragraph{\textbf{Implementation Details}}
We employ Qwen-Plus~\cite{qwen-plus} (2025-04-28) as the core LLM of our system for user intent parsing, tool planning, and the construction of offline conversational datasets. It supports a maximum input length of 126K tokens. 
The framework of ScienceDB AI is implemented using Python and LangGraph~\cite{langgraph}. We adopt a distributed Qdrant~\cite{qdrant} cluster as our online vector database. 
For dialogue-based models (e.g., DeepSeek and Qwen), we first use Approximate Nearest Neighbor (ANN) search to retrieve candidate datasets based on the researcher's query (as the tool results shown in Fig.~\ref{fig:dataset-sample-1}). The candidates are selected from over 10 million datasets in ScienceDB. These retrieved datasets are then provided as context to dialogue LLMs, which selects the final recommendation. All comparative baselines are conducted with their default hyper-parameters. For models that do not support multi-turn interactions (e.g., DataFinder and DataLinking), we decompose the multi-turn queries into a series of single-turn queries.
For our framework, we set $N$ to 30 as the default.

\subsection{Overall Performance}
\label{sec:exp_overall}
We first evaluate the overall performance of ScienceDB AI and its competitors in our offline multi-turn conversational recommendations. The results are shown in Table \ref{tab:main_res}. We summarize our key findings as follows: \textbf{(1)}  Existing models specifically designed for dataset recommendation (DataFinder and DataLinking), perform poorly. These models primarily rely on shallow semantic similarity between input queries and dataset descriptions, making them inadequate for understanding the complicated and domain-specific needs of researchers. Notably, DataFinder shows particularly poor performance due to its reliance on simple keyword-based similarity.
 \textbf{(2)} Agent-based models outperform dialogue-based LLMs, demonstrating the effectiveness of incorporating agent structures.
\textbf{(3)} Our proposed ScenceDB.AI consistently outperforms all competitors across all evaluation metrics, validating the effectiveness of our framework. Compared to the strongest baseline, InteRecAgent, ScenceDB.AI achieves more than a 20\% improvement. While InteRecAgent incorporates additional modules (e.g., the complicated planning and reflection module), it still underperforms relative to our more compact and efficient design.
\textbf{(4)} Based on the results of AT, we conclude that ScenceDB.AI has the smallest turn to find the true answer. Compared with the best AT competitor, Qwen, ScenceDB.AI has achieves 8\% and 10\% improvement in AT@3 and AT@5.
\textbf{(5)} We observe that most baseline models benefit significantly from increasing the value of $k$. For example, CoSearchAgent improves its Recall by 48.4\% from @1 to @5. In contrast, ScienceDB AI shows only a modest 3.2\% gain, as it already achieves high recall at top positions, reflecting its ability to rank the correct dataset near the top with high initial precision.

\subsection{Running Efficiency}
\begin{figure}[htbp]%
    \centering { \includegraphics[width=\linewidth]{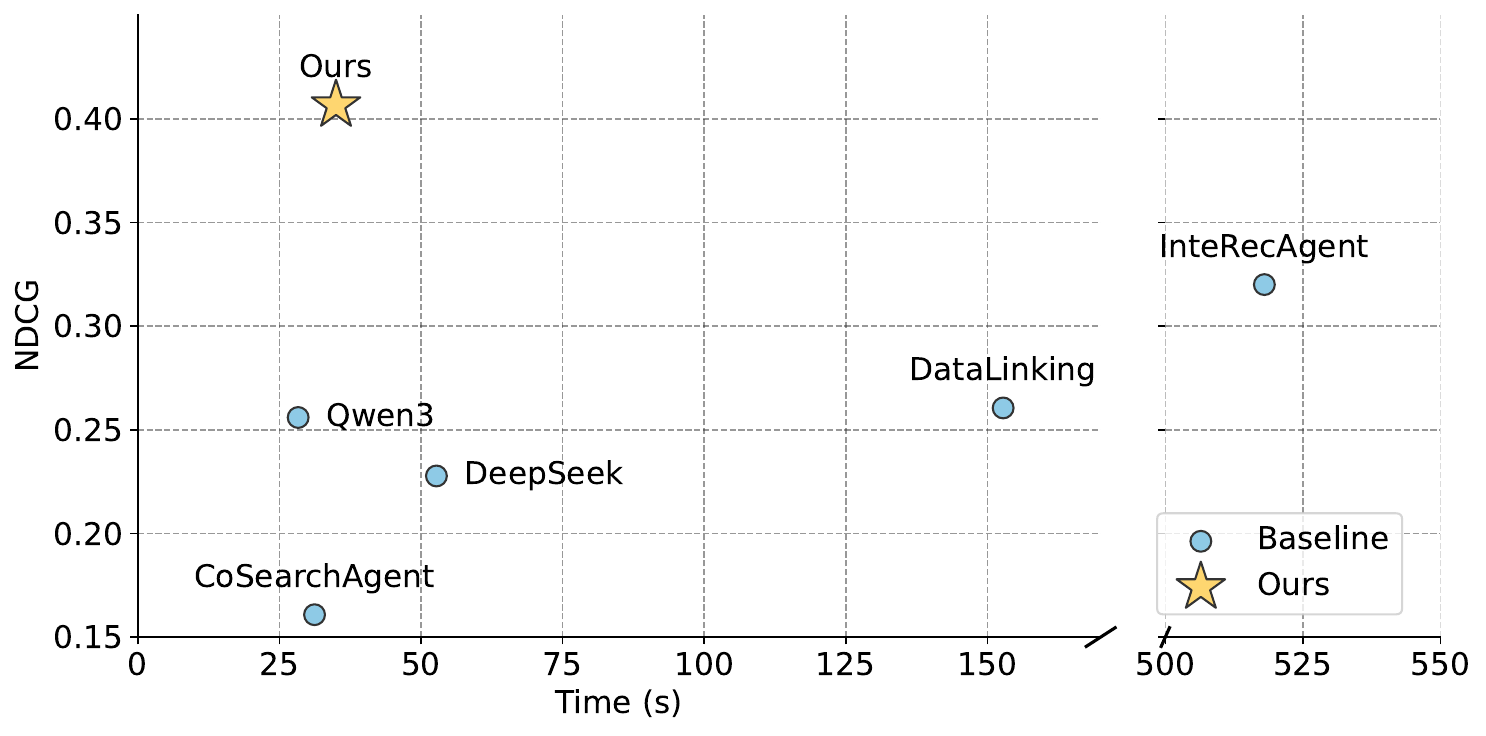} }
    \vspace{-3mm}
    \caption{The average running time for each testing sample.} 
    \label{fig:Efficiency}
    \vspace{-2mm}
\end{figure}

We evaluate the running efficiency of ScienceDB AI in comparison with other LLM- and agent-based conversational baselines. Fig.~\ref{fig:Efficiency} reports the average inference time per offline conversational test sample. As shown in the figure, InteRecAgent, which incorporates a self-reflection module, exhibits significantly higher inference time (518s) than all other models. Despite being a single-agent model, InteRecAgent runs slower than the multi-agent-based CoSearchAgent, highlighting the computational cost introduced by self-reflection.
Surprisingly, DataLinking, though based on simple keyword similarity rather than LLMs, still incurs longer inference time than several LLM-based approaches, indicating inefficiencies in its implementation. In contrast, ScienceDB AI demonstrates superior in both inference efficiency and effectivenss, making it highly practical for deployment in real-world, large-scale data sharing services.

\begin{figure}[htbp]
    \centering
    \includegraphics[width=\linewidth]{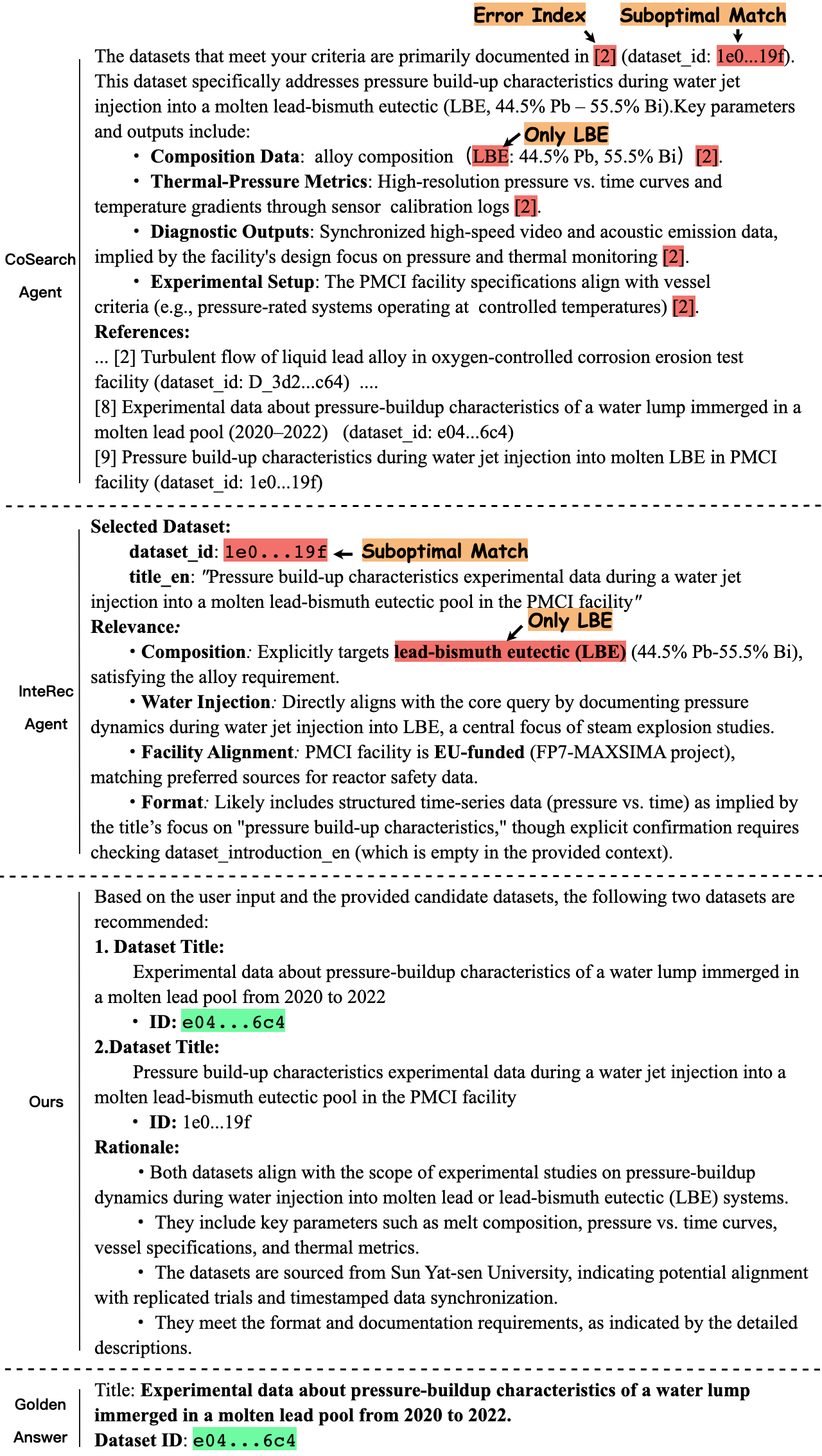}
    \caption{A case study of ScienceDB AI and its two competitive Agent-based recommenders.}
    \vspace{-3mm}
    \label{fig:case_study}
\end{figure}

\subsection{Case Study}
To effectively compare the performance, we present case studies in Fig.~\ref{fig:case_study}. We compare the outputs of two Agent-based recommenders, i.e., InteRecAgent and CoSearchAgent, and ScienceDB AI for a given experiment-level query. The input query is shown Fig.\ref{fig:dataset-sample-1} in the Appendix.
Specifically, when a researcher requests datasets on pressure-buildup dynamics during water injection into molten lead-bismuth alloys. The request includes eutectic alloys (44.5\% Pb–55.5\% Bi), non-eutectic compositions, and pure bismuth. The user also specifies the need for synchronized diagnostic outputs and stratified thermal conditions. Both the InteRec Agent and CoSearch Agent return the PMCI dataset. This dataset includes eutectic LBE experiments with pressure and temperature measurements. However, it fails to meet several key requirements: it only covers eutectic compositions and lacks data on non-eutectic and pure-metal cases. In addition to semantic mismatches, CoSearch also exhibits structural errors. For example, it mislabels dataset enumeration numbers, causing mismatches between dataset IDs and their corresponding descriptions. In contrast, our ScienceDB AI correctly identifies a more appropriate dataset. This dataset features high-resolution pressure traces from pure lead experiments conducted between 2020 and 2022, synchronized acoustic and video diagnostics, and comprehensive metadata with full documentation.

\subsection{Online A/B Test}
\begin{figure}[htbp]%
    \centering { \includegraphics[width=\linewidth]{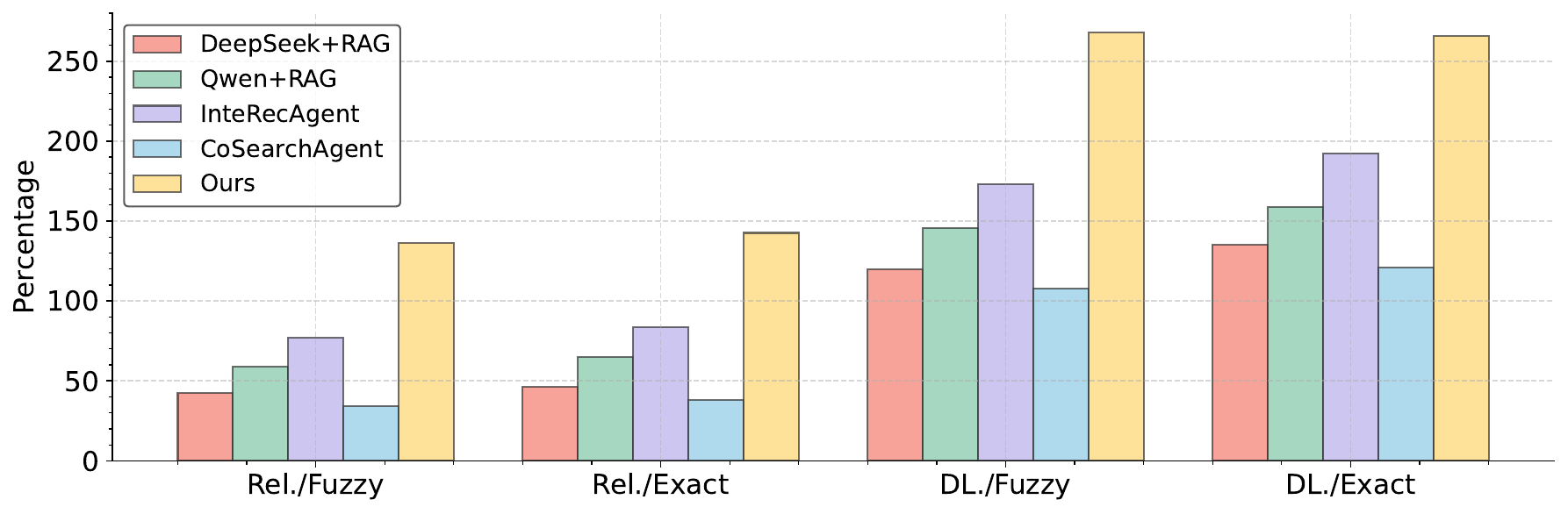} }
    \vspace{-3mm}
    \caption{Performance improvement of our ScienceDB AI over the original retrieval system in the online A/B test.} 
    \vspace{-2mm}
    \label{fig:online_res}
\end{figure}

ScienceDB AI introduces a new search interface to the original ScienceDB platform, we compare the CTR of ScienceDB AI and its competitive baselines with the original online keyword-based search page at \url{https://scidb.cn/en/list?searchList}, focusing on Top-4 positions. The baseline system includes four retrieval configurations: (1) relevance-based with fuzzy matching (Rel./Fuzzy), (2) relevance-based with exact matching (Rel./Exact), (3) download-frequency-based with fuzzy matching (DL./Fuzzy), and (4) download-frequency-based with exact matching (DL./Exact).  As shown in Fig.~\ref{fig:online_res}, all values indicate the relative improvements of our model and comparable baselines over the keyword-based search system, measured in percentage terms. We have the following findings: \textbf{(1)} ScienceDB AI achieves significantly higher CTRs, outperforming all baselines across all settings. The conclusion is consistent with the offline experiments in Table \ref{tab:main_res}. Notably, the improvement is more pronounced under exact matching conditions.  \textbf{(2)}  The performance improvements of Rel- and DL-based matching show no significant difference between the fuzzy and exact settings. The result indicates that traditional keyword-based dataset search methods fail to capture the semantics of input queries. Instead they lie in string matching, whether through exact matches requiring full identity or fuzzy matches based on character similarity, neither approach understands researchers’ scientific intentions.

\section{Conclusion}
In this paper, we introduced ScienceDB AI, an intelligent agentic recommender system for large-scale scientific data sharing, built on a repository of over 10 million high-quality scientific datasets. The system introduces several innovations: a Experimental Intention Perceptor to extract structured experimental elements from complicated queries, a Structured Memory Compressor to manage multi-turn dialogues effectively, and a Trustworthy Retrieval-Augmented Generation (Trustworthy RAG) framework. The Trustworthy RAG employs a two-stage retrieval mechanism and provides citable dataset references via Citable Scientific Task Record (CSTR) identifiers, enhancing recommendation trustworthiness and reproducibility. Through extensive offline and online experiments using large-scale real-world datasets, ScienceDB AI has demonstrated significant effectiveness, achieving about 30\% improvement in offline metrics compared to advanced baselines and a over 200\% increase in click-through rates compared to keyword-based search engines. To the best of our knowledge, ScienceDB AI is the first LLM-driven conversational recommender tailored explicitly for large-scale scientific dataset sharing services.

\bibliographystyle{ACM-Reference-Format}
\bibliography{cite}

\clearpage
\appendix

\section{Appendix}

\subsection{Detailed Offline Dataset Information}
\label{sec:Appendix_Offline_Dataset}
\paragraph{Source Data}
Fig.~\ref{fig:qdrant-sample} presents a representative dataset entry with typical structured metadata, including title, authorship, taxonomy classification, keywords, and a textual description. Such entries capture essential information for indexing and retrieval, and serve as the foundation for downstream tasks like dataset recommendation and semantic understanding.

\definecolor{lightgray}{gray}{0.95}
\begin{figure}[htbp]
\centering
\small
\begin{tcolorbox}[colback=gray!5!white, colframe=gray!80!black, title=A candidate dataset sample.]
\textbf{"title"}: "Experimental data about pressure-buildup characteristics of a water lump immerged in a molten lead pool from 2020 to 2022",\\
\textbf{"cstr"}: "31253.11.sciencedb.j00186.00022",\\
\textbf{"dataSetPublishDate"}: "2023-02-24T06:52:19Z", \\
\textbf{"author"}: [\\
\hspace*{2em}\{"name": "...", "organizations": [ "..." ] \},\\
\hspace*{2em}\{"name": "...", "organizations": [ "..." ]\} ], \\
\textbf{"taxonomy"}: [{"code": "490","nameZh": "","nameEn": "Nuclear science and technology"}], \\
\textbf{"keywordEn"}: ["Lead-cooled fast reactor","Steam generator tube rupture accident","Pressure-buildup characteristics","Experimental study"], \\
\textbf{"introduction"}: "To understand the pressure-buildup characteristics of a water droplet immerged inside a molten lead pool, which is a key phenomenon during a Steam Generator Tube Rupture accident of Lead-cooled Fast Reactor, many experiments have been conducted by injecting water lumps into a molten lead pool at Sun Yat-sen University from 2020 to 2022. In order to deepen the understanding of the influence of melt material, this lead experiment was compared with a Lead-Bismuth-Eutectic (LBE) experiment in the literature. The parameters employed in the experiments are water volume, water shape, water subcooling, molten pool depth and melt temperature.The interaction vessel in which the CCI occurs is a stainless steel cylindrical container with an inner diameter of 250 mm, a height of 750 mm, and a design pressure of 40 MPa. Many sensors are installed on the interaction vessel wall to obtain the temperature and pressure trends of the melt pool and cover gas."\\

\end{tcolorbox}
\caption{A candidate dataset sample, which containing metadata and descriptions.}
\label{fig:qdrant-sample}
\end{figure}

\paragraph{Offline Evaluation Dataset Construction Pseudocode}
Algorithm~\ref{alg:offline_dataset_process} outlines the procedure for constructing a simulated multi-turn conversation entry $e$ based on a user’s historical interactions. Given a user ID $u$, a sequence of historical items $H = [h_1, h_2, ..., h_n]$, a selected target index $i$, a template module $T$, and the maximum number of interaction rounds $R$, the algorithm generates a synthetic dialogue that reflects a realistic yet challenging information-seeking process.
In \textbf{Step 1}, the algorithm selects a fixed-length history window $H_{\text{sel}} = H[i{-}L:i]$ preceding the target index $i$. The target item $d = H[i]$ represents the dataset the user truly intends to retrieve. A new conversation entry $e$ is initialized using $u$, $i$, $T$, and $H_{\text{sel}}$. The titles of items in $H_{\text{sel}}$ are concatenated into a string $s$, which is passed to the system prompt generator $T.\texttt{sys\_prompt}(s)$ and appended to the entry.
In \textbf{Step 2}, a camouflaged user query $q_0$ and its associated supervision mask $\text{mask}$ are generated by \texttt{GenFakeRequest} based on $T$, $s$, and the ground-truth dataset $d$. This query is designed to indirectly express the user’s true intent The mask is stored in $e$, and the query is formatted using $T.\texttt{fake\_request}(q_0)$ before being appended as a user message.
\textbf{Step 3} simulates the multi-turn conversation loop. In each round $r$, the latest user and assistant messages $(u_r, a_r)$ are extracted from $e$, and a new tool query $q_r$ is constructed using $T.\texttt{tool\_query}(s, a_r, u_r)$. If $q_r$ contains a predefined end-of-task marker, the loop terminates. Otherwise, the query is sent to a retrieval backend via \texttt{Search}, and the returned documents are formatted as $R_r$. The assistant then generates a response $a_{r+1} = T.\texttt{generate\_response}(s, a_r, u_r, R_r)$, followed by a user follow-up $u_{r+1} = T.\texttt{user\_followup}(s, d, a_{r+1})$. Both messages are appended to $e$, and the loop halts early if $u_{r+1}$ also contains an end marker.
Finally, in \textbf{Step 4}, the ground-truth answer corresponding to the target dataset $d$ is generated via $T.\texttt{truth\_response}(d)$ and appended to the conversation as the assistant’s final reply.

\begin{algorithm}[ht]
\small
\caption{Generate Multi-turn Conversation}
\label{alg:offline_dataset_process}
\KwIn{User ID $u$, History $H = [h_1, h_2, ..., h_n]$, Target Index $i$, Template $T$, Max Rounds $R$}
\KwOut{Conversation entry $e$}

\vspace{1mm}
\tcp{\textbf{Step 1: Initialize entry}}
$H_{\text{sel}} \gets H[i{-}L : i]$\;
$d \gets H[i]$ \tcp*[r]{Target dataset}
$e \gets$ InitEntry($u$, $i$, $T$, $H_{\text{sel}}$)\;
$s \gets$ JoinTitles($H_{\text{sel}}$)\;
Append $(\texttt{System Prompt}, T.\texttt{sys\_prompt}(s))$ to $e$\;

\vspace{1mm}
\tcp{\textbf{Step 2: Generate initial user query}}
$(q_0, \text{mask}) \gets \texttt{GenFakeRequest}(T, s, d)$\;
$e.\text{mask} \gets \text{mask}$\;
Append $(\texttt{User Request}, T.\texttt{fake\_request}(q_0))$ to $e$\;

\vspace{1mm}
\tcp{\textbf{Step 3: Multi-turn simulation}}
\For{$r = 1$ \KwTo $R{-}1$}{
    $(u_r, a_r) \gets$ LastMessages($e$)\;

    $q_r \gets T.\texttt{tool\_query}(s, a_r, u_r)$\;
    Append $(\texttt{Invoke Tool}, q_r)$ to $e$\;

    \If{EndMarker($q_r$)}{
        \textbf{break}\;
    }

    $R_r \gets \texttt{Search}(q_r)$\;
    Append $(\texttt{Tool Result}, \texttt{Json}(R_r))$ to $e$\;

    $a_{r+1} \gets T.\texttt{generate\_response}(s, a_r, u_r, R_r)$\;
    Append $(\texttt{Assistant Response}, a_{r+1})$ to $e$\;

    $u_{r+1} \gets T.\texttt{user\_followup}(s, d, a_{r+1})$\;
    Append $(\texttt{User Request}, u_{r+1})$ to $e$\;

    \If{EndMarker($u_{r+1}$)}{
        \textbf{break}\;
    }
}

\vspace{1mm}
\tcp{\textbf{Step 4: Final ground-truth answer}}
Append $(\texttt{Final Answer}, T.\texttt{truth\_response}(d))$ to $e$\;

\KwRet{$e$}\;
\end{algorithm}

\definecolor{lightgray}{gray}{0.95}
\begin{figure*}[htbp]
\centering
\small
\begin{tcolorbox}[colback=gray!5!white, colframe=gray!80!black, title=A offline simulated conversational dataset recommendation sample.]
\{ \\
\hspace*{2em}\textbf{"action\_type"}:\textbf{"Simulated User Request"},\\
\hspace*{2em}\textbf{"content"}: "\\
\hspace*{4em}\underline{\textbf{Dataset Scope}}: Experimental datasets documenting pressure-buildup dynamics during water injection into molten lead-bismuth alloys (including eutectic LBE: 44.5\% Pb-55.5\% Bi, non-eutectic ratios like 20\% Pb-80\% Bi or 80\% Pb-20\% Bi) or pure bismuth. Studies must report under both stratified (e.g., 400°C bottom, 500°C top) and uniform melt temperatures ($\sim$500°C).  Key Parameters:\\
\hspace*{4em}1. \underline{\textbf{Composition Data}}: Full alloy composition (\% mass of Pb/Bi), melt material properties (density, thermal conductivity, viscosity).\\
\hspace*{4em}2. \underline{\textbf{Thermal-Pressure Metrics}}: High-resolution pressure vs. time curves, temperature gradients with spatial resolution (e.g., axial thermocouples), water injection parameters (50 mL volume, 80°C subcooled water, droplet morphology).\\
\hspace*{4em}3. \underline{\textbf{Diagnostic Outputs}}: Synchronized high-speed video (fragmentation modes, vapor-layer collapse) and acoustic emissions (frequency spectra, amplitude bursts tied to pressure spikes), with metadata linking precursor signals (e.g., vapor collapse) to pressure kinetics.\\
\hspace*{4em}4. \underline{\textbf{Experimental Setup}}: Vessel specifications (e.g., stainless steel, 250 mm ID × 750 mm height, 40 MPa-rated), sensor calibration logs for pressure/temperature, melt preparation/injection protocols (induction heating, thermocouple arrays).  Exclusions: - Non-metallic melts or non-water coolants (e.g., sodium, CO2).  Preferred Sources:  - OECD/NEA databases, IAEA reactor safety programs, FP7-MAXSIMA datasets, or other EU/NRC-funded experiments on steam explosion physics.  - Replicated trials ($\geq$5 per condition) with timestamps aligning pressure, thermal, and diagnostic data.  - Comparative datasets for pure Pb vs. bismuth-rich alloys.  Format Requirements:  - Structured time-series data (CSV, HDF5, MATLAB .mat, Excel).  - Timestamped video (AVI, MP4) with metadata.  - Documentation: README files, sensor calibration records, experimental schematics.  Search Keywords: Molten metal-water interaction, steam explosion pressure dynamics, lead-bismuth alloy thermal stratification, bismuth coolant safety experiments, acoustic precursors in vapor explosions, reactor cooling system fragmentation studies.  Purpose: Benchmark peak pressure values, pressure rise rates, and acoustic precursors; validate diagnostic synchronization (e.g., video-acoustic alignment) against prior experimental data." \\
\}, \\
\{ \\
\hspace*{2em}\textbf{"action\_type"}: \textbf{"Tool Result"},\\
\hspace*{2em}\textbf{"content"}: "[\\
\hspace*{4em}[0.7737, {\textbf{dataset\_id: e04...6c4}, title: Experimental data about pressure-buildup characteristics of a water lump immerged in a molten lead pool from 2020 to 2022., dataset\_introduction: ... }], \\
\hspace*{4em}[0.7697, {\textbf{dataset\_id: 1e0...19f}, title: Pressure build-up characteristics experimental data during a water jet injection into a molten lead-bismuth eutectic pool in the PMCI facility, dataset\_introduction: ...}],\\
\hspace*{4em}..., \\
\hspace*{4em}[0.6674, {\textbf{dataset\_id: D\_3d2...c64}, title: Turbulent flow of liquid lead alloy in oxygen-controlled corrosion erosion test facility, dataset\_introduction: ...}], \\
\hspace*{4em}..." \\
\}, \\
\{ \\
\hspace*{2em}\textbf{"action\_type"}: \textbf{"Model Response"},\\
\hspace*{2em}\textbf{"content"}: "......" \\
\}, \\
\{ \\
\hspace*{2em}\textbf{"action\_type"}: \textbf{"True Answer"},\\
\hspace*{2em}\textbf{"content"}: "According to your request, here is the dataset you wanted: Experimental data about pressure-buildup characteristics of a water lump immerged in a molten lead pool from 2020 to 2022. The dataset id is: e04...6c4." \\
\} \\
\end{tcolorbox}
\vspace{-2mm}
\caption{A sample of our simulated offline conversational dataset recommendation.}
\label{fig:dataset-sample-1}
\end{figure*}

\paragraph{Specific Dataset Instances/Samples}
We provide a specific offline dataset instance, which is shown in Fig.~\ref{fig:dataset-sample-1}.
Each dataset sample consists of a simulated query of researchers, tool results, the response of model outputs, and true answer. We construct simulated user queries by modeling detailed experimental requirements, such as Composition Data, Thermal-Pressure Metrics, Diagnostic Outputs, and Experimental Setup. This method captures realistic user needs more comprehensively than simple keyword queries, enabling better evaluation of dataset recommendation systems. Based on this request, the system retrieves candidate datasets from a structured database this is recorded as the tool result (candidate datasets). Finally, the true answer provides the dataset the simulated user was actually intended to find.

\subsection{Detailed Information of Evaluation Metrics}
\label{appendix:llm-metrics}
To assess the performance of the dataset recommendation models, we adopt the following widely used and representative ranking based metrics: Recall@k, NDCG@k, and MRR@k, with \( k \in \{1, 3, 5\} \). 
These metrics are defined as follows:

\paragraph{Recall@k}
Recall@k measures whether the ground-truth relevant item is ranked within the top-$k$ positions:
\[
\text{Recall@}k = 
\begin{cases}
1, & \text{if relevant item is ranked } \leq k. \\
0, & \text{otherwise}.
\end{cases}
\]
In our setting, each query has a single relevant dataset, so Recall@k evaluates the hit rate at position $k$.

\paragraph{NDCG@k (Normalized Discounted Cumulative Gain)}
NDCG@k considers the position of the relevant item in the ranked list, assigning higher weights to items ranked higher. It is defined as:
\[
\text{NDCG@}k = \frac{1}{\log_2(r+1)} \quad \text{if relevant item is at rank } r \leq k
\]
Otherwise, NDCG@k = 0. When there is only one relevant item, the ideal DCG (IDCG@k) is 1, so NDCG@k simplifies to a single-position discount.

\paragraph{MRR@k (Mean Reciprocal Rank)}
MRR@k measures the inverse of the rank position of the first relevant item, truncated at $k$:
\[
\text{MRR@}k = 
\begin{cases}
\frac{1}{r}, & \text{if relevant item is at rank } r \leq k. \\
0, & \text{otherwise}.
\end{cases}
\]
We report the average MRR@k over all queries. All metrics are averaged over the test set and evaluated at $k = 1, 3, 5$ to assess ranking quality at various depths.

To further evaluate recommendation efficiency in multi-turn dialogues, we propose a new metric: \textbf{AT (Average Turn)}. This metric measures how early the model is able to recommend the correct dataset within a conversation.
Formally, for each multi-turn dialogue, we identify the first turn $t$ in which the model's response includes the ground-truth dataset in the top-$k$ results. The AT score for that dialogue is defined as:
\[
\text{AT} = 
\begin{cases}
t, & \text{if the correct dataset appears in turn } t \leq T \\
T + 2, & \text{if the correct dataset is not found in any turn}
\end{cases}
\]
where $T$ denotes the total number of dialogue turns. The penalty of $T + 2$ ensures that dialogues where the model fails entirely are appropriately penalized.
The final AT score is computed as the average over all dialogues.

\paragraph{Metric Extension for Multi-turn Dialogues}

In contrast to traditional single-turn settings, our dataset features multi-turn conversational queries where users iteratively refine their requests. To reflect this process, we adopt a global top-$k$ evaluation strategy: instead of averaging metrics (Recall@k, NDCG@k, MRR@k) over individual turns, we concatenate model responses in reverse chronological order (from last to first turn) and compute metrics on the resulting ranked list, prioritizing later, more specific intents.

However, standard metrics do not distinguish whether the correct dataset is identified early or late. To capture interaction efficiency, we propose AT to reflect the earliest turn at which the correct dataset is recommended. A lower AT indicates quicker task resolution and better understanding. By combining AT with standard metrics, we provide a more holistic evaluation of both recommendation quality and efficiency.

\end{document}